\documentclass[sigconf, screen, nonacm]{acmart}
\usepackage{ulem}
\usepackage{multirow}

\AtBeginDocument{%
  }

\setcopyright{acmlicensed}
\copyrightyear{2025}
\acmYear{2025}
\acmDOI{XXXXXXX.XXXXXXX}
\acmConference[Multimedia '25]{Make sure to enter the correct
  conference title from your rights confirmation email}{June 27--31,
  2025}{Dublin, Ireland}
\acmISBN{978-1-4503-XXXX-X/2018/06}

\begin{document}

\title{FlexSpeech: Towards Stable, Controllable and Expressive Text-to-Speech}

\author{Linhan Ma}
\authornote{Both authors contributed equally to this research.}
\email{mlh2023@mail.nwpu.edu.cn}
\affiliation{%
  \institution{Northwestern Polytechnical University}
  \city{}
  \country{}
}
\author{Dake Guo}
\authornotemark[1]
\email{guodake@mail.nwpu.edu.cn}
\affiliation{%
  \institution{Northwestern Polytechnical University}
  \city{}
  \country{}
}

\author{He Wang}
\email{hwang2001@mail.nwpu.edu.cn}
\affiliation{%
  \institution{Northwestern Polytechnical University}
  \city{}
  \country{}
}

\author{Jin Xu}
\authornote{Corresponding Authors.}
\affiliation{%
  \institution{Independent Researcher}
  \country{}
}

\author{Lei Xie}
\authornotemark[2]
\email{lxie@nwpu.edu.cn}
\affiliation{%
  \institution{Northwestern Polytechnical University}
  \city{}
  \country{}
}


\begin{abstract}
Current speech generation research can be categorized into two primary classes: non-autoregressive~(NAR) and autoregressive~(AR). The fundamental distinction between these approaches lies in the duration prediction strategy employed for predictable-length sequences. The NAR methods ensure stability in speech generation by explicitly and independently modeling the duration of each phonetic unit. Conversely, AR methods employ an autoregressive paradigm to predict the compressed speech token by implicitly modeling duration with Markov properties. Although this approach improves prosody, it does not provide the structural guarantees necessary for stability. To simultaneously address the issues of stability and naturalness in speech generation, we propose \textbf{FlexSpeech}~\footnote{FlexSpeech means our TTS system is flexible and controllable in speech geeneration and can effortlessly transfer specific style to unseen speakers using a very small amount of data.}, a stable, controllable, and expressive TTS model. The motivation behind FlexSpeech is to incorporate Markov dependencies and preference optimization directly on the duration predictor to boost its naturalness while maintaining explicit modeling of the phonetic units to ensure stability. Specifically, we decompose the speech generation task into two components: an AR duration predictor and a NAR acoustic model. The acoustic model is trained on a substantial amount of data to learn to render audio more stably, given reference audio prosody and phone durations. The duration predictor is optimized in a lightweight manner for different stylistic variations, thereby enabling rapid style transfer while maintaining a decoupled relationship with the specified speaker timbre.
Experimental results demonstrate that our approach achieves SOTA stability and naturalness in zero-shot TTS. More importantly, when transferring to a specific stylistic domain, we can accomplish lightweight optimization of the duration module solely with about 100 data samples, without the need to adjust the acoustic model, thereby enabling rapid and stable style transfer. Audio samples can be found in our demo page~\footnote{\href{https://flexspeech.github.io/DEMO/}{https://flexspeech.github.io/DEMO/}}.

\end{abstract}





\maketitle

\section{Introduction}

The progression of neural text-to-speech (TTS) systems has been propelled by alternating advancements in autoregressive and non-autoregressive speech generation methodologies, resulting in a spiral evolution within the field. Early works, such as Tacotron~\cite{tacotron, tacotron2} and FastSpeech~\cite{fastspeech, fastspeech2}, have achieved state-of-the-art (SOTA) performance, alternating between naturalness and stability.  As general artificial intelligence~(AGI) technologies continue to advance rapidly, the standards for speech generation models increasingly target the attainment of human-level naturalness. Recently, discrete speech representation codecs~\cite{soundstream, encodec, hificodec, speechtokenizer, DACcodec, fsq} have gained prominence in the field. This line of research involves training a codec to identify appropriate discrete speech compression units, which are then modeled in an end-to-end autoregressive manner. By implicitly modeling the Markov dependencies among various phonetic units, the naturalness of the generated speech has significantly improved, achieving a level that is comparable to human performance.

Despite the impressive results achieved with end-to-end codec modeling based on autoregressive (AR) models~\cite{BaseTTS,valle,valle2,seedtts,tortoise}, these approaches continue to suffer from instability due to a lack of architectural guarantees. For instance, when generating audio from complex, long sentences, existing AR models are prone to issues such as repetition and word omission. Additionally, AR models are inherently susceptible to cascading generation failures when confronted with unknown tokens and their combinations. In contrast, some non-autoregressive (NAR) models~\cite{ns2,ns3,flshspeech,mega2}, ensure system stability by explicitly modeling the duration of phonetic units. However, most approaches do not account for the dependencies between duration units, leading to audio output that lacks richness in prosody and rhythm.

The motivation for this work is to enhance the naturalness and stylistic transfer capabilities of speech generation while ensuring stability by integrating Markov dependencies and preference optimization relationships into duration modeling. To tackle these challenges simultaneously, we propose FlexSpeech, a text-to-speech (TTS) model characterized by its stability, controllability, and expressiveness. 
In our approach, we decompose the speech generation task into two distinct components: the autoregressive duration predictor and the non-autoregressive acoustic model. 
The acoustic model, founded on flow matching~\cite{flowmatching}, is trained on a comprehensive dataset to enhance audio rendering based on the prosody of reference audio and phoneme durations, directly predicting mel-spectrograms from reference speaker features and phoneme sequences with integrated duration information. Meanwhile, the duration model employs an encoder-decoder architecture that incorporates reference acoustic representations and phoneme-duration prompts to predict target phoneme durations in an autoregressive manner. For each phoneme, a discrete label is defined as the corresponding frame number of the mel-spectrogram and optimized using cross-entropy loss, with the prediction for subsequent phoneme durations performed via next-token prediction. This modular design preserves the advantages of AR models in capturing sequential dependencies and expressive prosody while ensuring synthesis stability through accurate duration modeling. Moreover, by incorporating Direct Preference Optimization (DPO)~\cite{dpo} with a modest set of win-lose duration pairs, the system aligns predicted durations with human auditory preferences, enabling efficient stylistic adaptation without the need to retrain the full acoustic model.

Our work is closely related to MegaTTS~\cite{mega,mega2,mega3}. The primary differences lie in two aspects. First, from a motivational perspective, we recognize that the existence of diverse data is justifiable. We train the duration predictor to learn various phonetic distributions and then employ DPO to conduct lightweight preference optimization, selecting generation paths that align with human preference. In contrast, MegaTTS does not implement a DPO mechanism.
Second, in terms of performance, FlexSpeech not only ensures stability and achieves superior naturalness but also demonstrates rapid and stable style transfer capabilities. Our work is also related to DPO-based autoregressive TTS models~\cite{SpeechAlign,UNO,RIO,Preference_ailmt}, which enhance model stability by optimizing the Word Error Rate (WER) through DPO. In contrast, FlexSpeech guarantees stability through its design mechanisms, where DPO is primarily utilized to refine the rhythm and naturalness of the phonetic outputs in the duration predictor.

Comprehensive experiments demonstrate that our approach achieves state-of-the-art stability with word error rate of 1.20\% in Seed-TTS \textit{test-zh} and 1.81\% in Seed-TTS \textit{test-en} and naturalness with best subjective evaluation scores. Impressively, we can realize rapid and stable speaking style transfer by lightweight optimization of the duration module solely with about 100 data samples without the need to adjust the acoustic model.

\begin{figure*}[htbp]
    \centering
    \includegraphics[width=\linewidth]{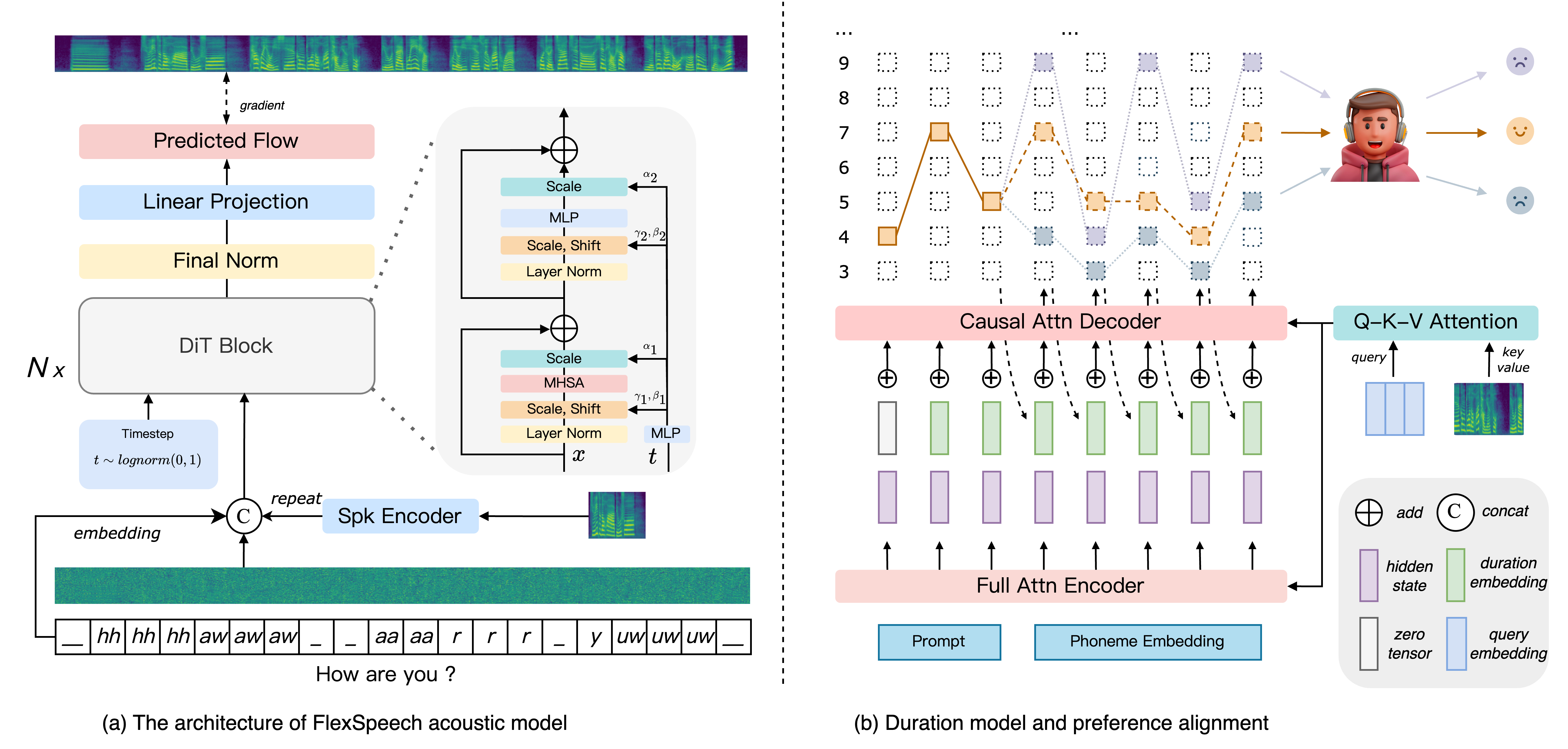}
    \caption{The overview of our FlexSpeech. (a). The architecture of FlexSpeech acoustic model. The phoneme embeddings, repeated speaker embeddings, and random noise are concatenated along the channel dimension to form the input to the model, which then predicts a mel-spectrogram of the same length. (b). Duration model and preference alignment. The decoder predicts current phoneme duration in an autoregressive manner, based on the hidden state and previous phoneme durations.}
    \label{system_overview_fig}
    \Description{The left sub-image shows the architecture of our flow-matching-based text-to-speech model. The right sub-image shows the architecture of our autoregressive duration model and the preference alignment process.}
\end{figure*}

\section{Preliminaries}

\subsection{Flow Matching}

Flow Matching (FM) aims to learn a probability path that transforms a complex data distribution \( p_t \) into a simpler one \( p_0 \), typically modeled as \( p_0 \sim \mathcal{N}(0,1) \). It shares similarities with Continuous Normalizing Flows (CNFs)~\cite{DBLP:conf/nips/ChenRBD18} but achieves significantly higher training efficiency through a simulation-free approach, reminiscent of the training paradigms used in diffusion probabilistic models (DPMs).

We can define the flow \( \phi: \mathbb{R}^d \to \mathbb{R}^d \) as the transformation that maps one density function to another, subject to the following ordinary differential equation:
\begingroup
\begin{align}
\label{fm:1}
d\phi _t (x) = v_t(\phi _t(x))dt, t \in [0,1]; \phi_0(x)=x, 
\end{align}
\endgroup

where \(v_t(\cdot)\) denotes the time-dependent vector field that defines the generation path \(p_t\) as the marginal probability distribution of the data points \(x\). Sampling from the approximate data distribution \(p_1\) is achieved by solving the initial value problem specified in the equation.

We assume a vector field \( u_t \) which generates a probability path \( p_t \) transitioning from \( p_0 \) to \( p_1 \). The FM loss is defined as:

\begingroup
\begin{align}
\label{fm:2}
\mathcal{L}_{\text{FM}}(\theta) = \mathbb{E}_{t,p_t(x)} ||  u_t(x)-v_t(x;\theta) ||^2,
\end{align}
\endgroup
where \( v_t(x; \theta) \) is a neural network parameterized by \( \theta \). However, implementing this approach is challenging in practice because obtaining the vector field \( u_t \) and the target probability distribution \( p_t \) is nontrivial. Consequently, the Conditional Flow Matching (CFM) loss can be defined as:

\begingroup
\begin{align}
\label{fm:3}
\mathcal{L}_{\text{CFM}}(\theta) = \mathbb{E}_{t,q(x_1),p_t(x|x_1)}||u_t(x|x_1)-v_t(x;\theta)||^2.
\end{align}
\endgroup

CFM replaces the intractable marginal probability density and vector field with their conditional counterparts. A key advantage is that these conditional densities and vector fields are readily available and have closed-form solutions. Moreover, it can be shown that the gradients of \( \mathcal{L}_{\text{CFM}}(\theta) \) and \( \mathcal{L}_{\text{FM}}(\theta) \) with respect to \( \theta \) are identical~\cite{flowmatching}.

Building on optimal transport principles, the optimal-transport conditional flow matching (OT-CFM) method refines CFM by facilitating particularly simple gradient computations, thereby enhancing its practical efficiency. The OT-CFM loss function is defined as:
\begingroup
\begin{align}
\label{fm:3}
\mathcal{L}_{\text{OT-CFM}}(\theta) = \mathbb{E}_{t,q(x_1),p_0(x_0)} \left\| u_t(\phi_t(x) \mid x_1) - v_t(\phi_t(x);\theta) \right\|^2,
\end{align}
\endgroup
where the OT-flow is given by
\[
\phi_t = (1-t)x_0 + t x_1,
\]
which represents the linear interpolation from \( x_0 \) to \( x_1 \); here, each data point \( x_1 \) is paired with a random sample \( x_0 \sim \mathcal{N}(0, I) \). Furthermore, the gradient vector field, whose expectation corresponds to the target function we aim to learn, is defined as
\[
u_t(\phi_t(x_0) \mid x_1) = x_1 - x_0.
\]
This vector field is linear, time-dependent, and depends solely on \( x_0 \) and \( x_1 \). These properties simplify the training process, enhance efficiency, and improve both generation speed and performance compared to diffusion probabilistic models (DPMs).

In our proposed approach, we transform a random sample \(x_0\) from the standard Gaussian noise to \(x_1\), the target mel-spectrogram, under the condition of corresponding aligned phoneme tokens and the speaker embedding from a reference mel-spectrogram. Hence, the final loss can be described as:
\begingroup
\begin{align}
\label{fm:3}
\mathcal{L}_{\text{FM}}(\theta)=\mathbb{E}_{t,q(x_1),p_0(x_0)}||(x_1-x_0)-v_t((1-t)x_0+tx_1,c;\theta)||^2
\end{align}
\endgroup
where \(c\) represents an additional condition, which is the concatenation of aligned phoneme tokens and the speaker embeddings.

Classifier-Free Guidance (CFG)~\cite{cfg} has been demonstrated to improve the generation quality of diffusion probabilistic models. It replaces an explicit classifier with an implicit one, eliminating the need to compute both the classifier and its gradient. The final generation result can be guided by randomly dropping the conditioning signal during training and performing linear extrapolation between inference outputs with and without the condition \( c \). 
During generation, the vector field is modified as follows:
\begingroup
\begin{align}
\label{cfg}
v_{t,\text{CFG}} = (1+\alpha)\cdot v_t(\phi _t(x),c;\theta)  - \alpha \cdot v_t(\phi _t(x);\theta)
\end{align}
\endgroup
where \(\alpha\) is the extrapolation coefficient of CFG.

\subsection{Preference Alignment}

Preference alignment is often formulated as a reinforcement learning problem. Let \( x \) denote the input prompts and \( y \) the corresponding response from the language model. Given a reward function \( r(x, y) \) and a reference policy \( \pi_{\text{ref}} \), the goal of alignment is to optimize the aligned policy \( \pi_\theta \) to maximize the expected reward while remaining close to the reference policy. This objective is expressed as:
\begingroup
\begin{align}
\label{eq:1}
\max_{\pi_\theta} \ \mathbb{E}_{y \sim \pi_\theta(y|p)} \left[ r(p, y) \right] 
- \beta \, D_{\text{KL}} \left( \pi_\theta(y|p) \,\Vert\, \pi_{\text{ref}}(y|p) \right),
\end{align}
\endgroup
where \( \beta \) is a hyperparameter that mediates the balance between maximizing the expected reward and penalizing deviations from the reference policy via the KL divergence term.

The KL-divergence term, regulated by the hyperparameter \(\beta\), prevents the aligned policy from deviating too far from the reference policy. A higher \(\beta\) imposes a stronger constraint. 
In practice, the reward function \( r \) is usually unknown and is inferred from human preference data in the form of tuples \((x, y_w, y_l)\), where \(y_w\) denotes the 'winner' (i.e., the preferred response) and \(y_l\) denotes the 'loser' (i.e., the disfavored response). Given such preference data, the reward function \(r\) can be estimated via maximum likelihood estimation:
\begingroup
\begin{align}
\hat{r} \in \arg\min_{r} \mathbb{E}_{(x, y_w, y_l)} \left[ -\log \sigma \Bigl( r(x, y_w) - r(x, y_l) \Bigr) \right],
\end{align}
\endgroup
where \(\sigma\) is the sigmoid function. With the estimated reward \(\hat{r}\), the policy \(\pi_\theta\) in Eq.~\ref{eq:1} can subsequently be optimized.

Notably, the optimization in Eq.~\ref{eq:1} can be solved in closed form without explicitly constructing a reward model. The direct preference optimization leverages the optimal solution to the KL-constrained objective to reparameterize the true reward function~\cite{dpo}. Specifically, the reward function is expressed as:
\begingroup
\begin{align}
r(x, y) = \beta \log \left( \frac{\pi_\theta(y|x)}{\pi_{\text{ref}}(y|x)} \right) + \beta \log Z(x),
\end{align}
\endgroup
where \(Z(x)\) is a normalization constant.

Under the Bradley-Terry model~\cite{Bradley-Terry}, the probability that \(y_w\) is preferred over \(y_l\) given input \(x\) is given by:
\begingroup
\begin{align}
P(y_w \succ y_l \mid x) = \sigma \left( \beta \log \left( \frac{\pi_\theta(y_w|x) \, \pi_{\text{ref}}(y_l|x)}{\pi_\theta(y_l|x) \, \pi_{\text{ref}}(y_w|x)} \right) \right).
\end{align}
\endgroup

Thus, the policy \(\pi_{\hat{\theta}}\) can be directly estimated from the preference data without an intermediate reward model. The DPO objective function is defined as:
\begingroup
\begin{align}
\label{eq:dpo}
\mathcal{L}_{\text{dpo}} = \mathbb{E}_{(y_w, y_l, x)} \left[ -\log \sigma \left( \beta \log \left( \frac{\pi_\theta(y_w|x) \, \pi_{\text{ref}}(y_l|x)}{\pi_\theta(y_l|x) \, \pi_{\text{ref}}(y_w|x)} \right) \right) \right].
\end{align}
\endgroup
The estimated policy is then obtained as:
\[
\pi_{\hat{\theta}}(y|x) \in \arg\min_{\pi_\theta} \mathcal{L}_{\text{dpo}},
\]
which implicitly maximizes the probability \(P(y_w \succ y_l \mid x)\).

\section{FlexSpeech}

In this section, we introduce our FlexSpeech, which is designed for speech synthesize with high expressiveness and naturalness that is in line with human preference. 
The overall architecture of our systems is shown in Figure~\ref{system_overview_fig}.
Our acoustic model takes a phoneme sequences with durations as input and directly predicts the corresponding mel-spectrogram. Then a vocoder, BigVGAN~\cite{bigvgan}, is employed to transform mel-spectrograms into waveforms. 
Phoneme durations are predicted by a separate duration model.
We separately pretrain and conduct supervised fine-tuning on these models via a large scale of data. At the last stage, to align the naturalness with human preference and style with target domain such as storytelling, we perform DPO on a few samples to efficient adjust the sampling pattern of duration models.

\subsection{Acoustic Model}
To begin with, the backbone of our acoustic model based on flow matching consists of Diffusion Transformer (DiT)~\cite{dit} blocks.
We also adopt zero-initialized adaptive LayerNorm (adaLN-zero) to enhance stability and controllability during training.
As shown in Figure~\ref{system_overview_fig} (a), we repeat each phoneme \(d_n\) times to obtain the length-expanded phoneme sequence, where \(d_n\) represents the duration of the \(n\)-th phoneme—specifically, the corresponding number of frames in the mel-spectrogram.
Consequently, the total length of the phoneme sequence exactly matches that of the mel-spectrogram, and we use this sequence directly as textual input for mel-spectrogram prediction. 
This approach eliminates the need for any padding or cropping operations, thereby significantly simplifying the modeling complexity of the acoustic system.
To achieve better decoupling and controllability, we employ reference speaker embedding for timbre modeling rather than an in-context learning approach like F5-TTS~\cite{f5tts}. The latter tends to incorporate the speaker’s duration characteristics, which may conflict with the duration inherent in the expanded phoneme sequence.
specifically, we utilize an ECAPA-TDNN-based~\cite{ecapa} speaker encoder module to extract an utterance-level speaker embedding vector from a reference mel-spectrogram random clip of several seconds.
Then, this vector is repeated to match the length of the expanded phoneme sequence and concatenates along the channel dimension with the phoneme embeddings and random noise, serving as the input to the DiT blocks.
We adopt logit-normal sampling instead of uniform sampling for timestep \(t\) to improve generation quality~\cite{sd3} and the timestep \(t\) is provided as the condition of adaLN-zero.

\subsection{Duration Model}
Given a finite-length phoneme sequence, duration modeling predicts each phoneme’s duration in a fixed-length autoregressive manner. As depicted in Figure~\ref{system_overview_fig}(b), our model utilizes an encoder-decoder architecture where each duration token is generated via next-token prediction. This methodology effectively captures the Markov dependencies inherent in natural speech, each prediction is conditioned on its immediate predecessor while simplifying the learning process through the decomposition of the joint probability distribution into a series of conditional probabilities.
The encoder consists of several transformer blocks with multi-head bidirectional attention. It takes phoneme embedding sequences as input and produces high-level hidden representations. To better encoder semantic information of the entire text, we add a mask learning loss to constrain the encoder. Specifically, we randomly select sentences during training and apply random masking to some of the phonemes within selected sentences. An additional linear project predicts the masked phonemes from the high-level hidden representations, and the cross-entropy loss function is employed as the constraint \(\mathcal{L}_{ml}\).

At each step \(n (n=1,2,...,N)\) where \(N\) is the length of the phoneme sequence, the hidden vector \(h_n\) is added with the previous phoneme's duration embedding \(e_{n-1}\), which is then fed into the decoder. 
At \(n=1\), we use a zero vector in place of duration embedding.
The decoder consists of several transformer blocks with multi-head causal attention to predict the duration label \(d_n\) and we also use the cross-entropy loss function.
Furthermore, to make better modeling, a reference mel-spectrogram clip is used as both the key and value in an attention mechanism, while a learnable embedding sequence is employed as the query. This operation yields an acoustically relevant feature \(feature_{ref}\) that is then fused into the hidden representation of both the encoder and decoder via cross-attention mechanisms.
The optimization objective of our duration model can be summarized as   
\begingroup
    \begin{align}
        \mathcal{L}_{ml} = &-\sum_{\hat{p}\in m(p)} \log p\left(\hat{p}|p_{\setminus m(p)} ; \theta_{enc}\right)
    \end{align}
\endgroup
\begingroup
    \begin{align}
        \mathcal{L}_{dur} = &-\sum_{n=0}^{N} \left(\log P\left(\overline{d}_{n} \mid \overline{d}_{<n}, \overline{{h}}_{\leq n}, {feature}_{ref} ; \theta_{dec}\right) \right)
    \end{align}
\endgroup
\begingroup
    \begin{align}
        \mathcal{L} = \lambda_{ml} \mathcal{L}_{ml} + \lambda_{dur}\mathcal{L}_{dur}
    \end{align}
\endgroup
where \(m(p)\) and \(p_{\setminus m(p)}\) denote the masked phonemes and the rest phonemes, \(\theta_{enc}\) and \(\theta_{dec}\) is the parameter of encoder and decoder, \(\lambda_{ml}\) and \(\lambda_{dur}\) are constant coefficients respectively.

\subsection{Inference Pipeline}
The in-context learning capability of our autoregressive duration model \(\theta_d\) enables it to predict phoneme-level durations that adhere to the same underlying patterns when provided with phoneme-duration prompts. 
In particular, the duration model employs the prompt durations \( d_{\text{prompt}} \), prompt phonemes \( p_{\text{prompt}} \), and a reference mel-spectrogram clip \( m_{ref} \) derived from the prompt, in addition to leveraging the historical context provided by all preceding durations \( d_{<n} \) and phonemes \( p_{\leq n} \) to predict target duration \(d_n\).
Then we use these to expand the target phoneme sequence and feed into the acoustic model with reference speaker embedding to generate a mel-spectrogram of equal length.
To sample from the learned distribution, the expanded phoneme sequence \(x_d\) and the reference speaker embedding \(se_{ref}\) serve as the condition in Eq.~\ref{cfg}
We have 
\begingroup
    \begin{align}
        v_t(\phi_t(x_0), c, \theta) = v_t((1-t)x_0 +  t x_1 | x_d, se_{ref})
    \end{align}
\endgroup
And the ODE solver is employed to integrate from \(\phi_0(x_0)=x_0\) to \(\phi_1(x_0)=x_1\) given \(d\phi_t(x_0)/dt = v_t(\phi_t(x_0), x_d, se_{ref}; \theta)\). After that, we use a BigVGAN vocoder to convert the predicted mel-spectrogram to 48kHz high-fidelity waveforms.

\subsection{Preference Alignment}

Within the autoregressive discrete duration prediction framework, we employ in-context learning to learn the prosody information in the prompt, which enables fine-grained control over phoneme durations. Even after SFT, the duration prediction model demonstrates robust generative capabilities and can statistically capture duration distributions effectively; however, our experiments reveal that the model tends to produce prosodic patterns that do not align with human preferences: for example, it may generate unnatural pauses or overly mechanical duration patterns, thereby compromising the naturalness of the synthesized speech. To mitigate this issue, we introduced a limited amount of manually annotated preference duration pairs and applied Direct Preference Optimization (DPO) to align the generated durations with human preferences. 
To differentiate between positive and negative audio samples, we provide annotators with detailed guidelines that emphasize critical aspects such as naturalness, abnormal pausing, and prosodic similarity. The annotation interface, as shown in Figure~\ref{interface}, is designed so that annotators only need to listen to each audio sample once to make their preference judgments.  Since annotators only need to listen to each audio sample once to make a preference judgment, the overall annotation process is highly efficient, achieving an average rate of 60 pairs per hour.
In our case, Eq.~\eqref{eq:dpo} can be modified to incorporate the duration-dependent components as follows:
\begingroup
\begin{align}
    \mathcal{L}_{\text{dpo-dur}} = \mathbb{E}_{(d_w, d_l, c)} \Biggl[ -\log \sigma \Biggl( \beta \log \Bigl( \frac{\pi_\theta(d_w \mid c) \, \pi_{\text{ref}}(d_l \mid c)}{\pi_\theta(d_l \mid c) \, \pi_{\text{ref}}(d_w \mid c)} \Bigr) \Biggr) \Biggr],
\end{align}
\endgroup
where \(d_w\) and \(d_l\) denote the ``winner'' and ``loser'' durations, respectively, and the conditioning variable\[ c = \Bigl( d_{\text{prompt}},\, p_{\text{prompt}},\, m_{ref} \Bigr)\]
This strategy preserves the benefits of autoregressive generation while ensuring that the predicted duration outputs more closely reflect the natural prosody patterns favored by humans, ultimately enhancing the overall naturalness and perceptual quality of the speech synthesis.

\section{Experiments and Results}

\subsection{Datasets}
We use the Emilia dataset, which is a multilingual and diverse in-the-wild speech dataset designed for large-scale speech synthesis, to pretrain our acoustic and duration model.
Only the English and Chinese data approximately 90K hours with valid transcriptions are retained for pre-training.
The Emilia dataset, due to being collected from the internet and auto-processed, contains background noise and transcription errors. This can lead to lower performance in sound quality and naturalness for the pretrained model. To address this, we applied supervised fine-tuning (SFT) using 1K hours of accurately annotated high-quality internal data to enhance the sound quality and naturalness.
We evaluate our zero-shot TTS system on three benchmarks: (1) LibriSpeech-PC \textit{test-clean} subset with 1127 samples in English released by F5TTS, (2) Seed-TTS~\cite{seedtts} \textit{test-en} with 1088 samples in English from Common Voice~\cite{commonvoice}, (3) Seed-TTS \textit{test-zh} with 2020 samples in Chinese from DiDiSpeech~\cite{didi}. During inference, we use the target speaker's phonemes and durations as prompts, with their mel-spectrogram providing a reference for timbre.

\subsection{Model Configuration}
We use 80-dimensional mel-spectrogram with a hop size of 160 and frame size of 1024 extracted from 16kHz downsampled speech for the acoustic and duration model training.
External alignment tool\footnote{https://github.com/MontrealCorpusTools/Montreal-Forced-Aligner} is used to extract the ground truth phoneme-level alignments.
A 22-layer DiT model is used as the backbone of our acoustic model with a hidden dimension of 1024, 16 attention heads, and a dropout rate of 0.1, with approximately 330M parameters. 
The speaker encoder utilizes ECAPA-TDNN architecture to extract 192-dimensional utterance-level speaker embeddings. As suggested in \cite{sd3}, logit-normal sampling is adopted instead of uniform sampling for timestep \(t\) to enhance generation quality.
During CFG training, conditions are dropped at a rate of 0.3. 
The acoustic model is pretrained on 8 
GPUs with a batch size of 30000 frames per GPU for 800K steps and then fine-tuned for 50k steps, and the AdamW optimizer with a peak learning rate of 9e-5 and 20K warm-up steps is used. 
For our duration model, the encoder and decoder each comprise 8 transformer layers that are interspersed with cross-attention layers. There are 512 hidden dimensions, 8 attention heads, and 0.1 dropout rate. All data samples that contain phoneme duration exceeding 99 are skipped. It is pretrained on 8 
GPUs with a batch size of 24 per GPU for 1M steps and then fine-tuned for 100K steps. 
Each sample has a probability of 0.5 to be selected, and within the selected samples, each phoneme has a probability of 0.15 to be masked.
\(\lambda_{ml}\) and \(\lambda_{dur}\) are 1.0 and 10.0 respectively.
The AdamW optimizer with a peak learning rate of 1e-4 and 20K warm-up steps is used.
For the vocoder, the original 1k-hour 48k audios are used to train the BigGAN from 16 kHz mel-spectrogram to 48kHz waveform. We follow the original configuration from BiGVGAN V2~\footnote{https://github.com/NVIDIA/BigVGAN} except for the upsample rates and upsample kernel sizes. In order to reconstruct a 48kHz waveform, the upsample rates are set to [5, 4, 3, 2, 2, 2] while the upsample kernel sizes are set to [11, 8, 7, 4, 4, 4]. BigVGAN is trained on 8 
GPUs with a batch size of 64 and a segment length of 48,000 for a total of 2M steps.
For inference, exponential moving average (EMA) weights are used for our acoustic model.
Euler solver is employed to compute the mel-spectrogram with 32-time steps and a CFG strength of 2.
We use a top-k of 6, top-p of 0.5, temperature of 0.9, and repetition penalty of 1.0 for duration model sampling.

\subsection{Evaluation Metrics}
For the objective metrics, we evaluate speaker similarity and intelligibility to demonstrate the stability of our system. Specifically, 
we compute the cosine similarity between speaker embeddings of generated samples and original references, which are extracted by a WavLM-large-based speaker verification model, for speaker similarity (SIM-O).
We utilize the Whisper-large-v3~\cite{whisper}~\footnote{https://huggingface.co/openai/whisper-large-v3} and the Paraformer-zh~\cite{para}~\footnote{https://huggingface.co/funasr/paraformer-zh} to transcribe English and Chinese samples respectively and then compute Word Error Rate (WER) for intelligibility.
For the subjective metrics, comparative mean opinion score (CMOS) is used to evaluate naturalness and expressiveness, and similarity mean opinion score (SMOS) is used to evaluate speaker similarity of timbre reconstruction and prosodic pattern. CMOS and SMOS are on a scale of -3 to 3 and 1 to 5, respectively.
We randomly select 15 samples from each test set for evaluation, ensuring that each sample is listened to by at least 10 individuals.

\begin{table*}[]
\renewcommand{\arraystretch}{0.85}
\small

\caption{Evaluation results on LibriSpeech \textit{test-clean}, LibriSpeech-PC \textit{test-clean}, Seed-TTS \textit{test-en} and \textit{test-zh}. The boldface and underline denote the best and second best result respectively. * means the results reported in original papers, ${\dagger}$ means we obtain the evaluation result using the official code and the pre-trained checkpoint, and $\diamondsuit$ means we download the samples from the demo page and inference our system with the same utterances to generate parallel samples. \#50 means the DPO utilizes 50 winner-loser data pairs.}

\begin{tabular}{ccccccc}
\hline
\textbf{Model}    & \textbf{\#Parameters} & \textbf{Training Data} & \textbf{CMOS↑}               & \textbf{WER(\%)↓} & \textbf{SIM-O↑}   & \textbf{SMOS↑}            \\ \hline
\multicolumn{7}{c}{\textbf{LibriSpeech \textit{test-clean}}}                                                                                                          \\ \hline
GT                & -                     & -                      & -                            & $2.2$             & $0.754$           & -                         \\
MaskGCT           & 1048M                 & 100kh Multi.           & $-0.25^{\dagger}$            & $2.643$*  & \uline{$0.687$}* & $3.91^{\dagger}$          \\
NaturalSpeech 3   & 500M                  & 60kh EN                & $-0.03^\diamondsuit$    & \textbf{$1.94$}*  & $0.67$*   & -                         \\
MegaTTS 2         & 300M+1.2B             & 60kh EN                & $-0.08^\diamondsuit$         & $2.73$*           & -                 & -                         \\ \hline
\multicolumn{7}{c}{\textbf{LibriSpeech-PC \textit{test-clean}}}                                                                                                       \\ \hline
GT                & -                     & -                      & $-0.19$                      & $2.23$            & $0.69$            & $3.93$                    \\
CosyVoice         & $\sim$300M            & 170kh Multi.           & $-0.49^{\dagger}$            & $3.59$*           & $0.66$*   & $3.82^{\dagger}$          \\
FireRedTTS        & $\sim$580M            & 248kh Multi.           & $-0.90^{\dagger}$            & $2.69$*           & $0.47$*           & $3.60^{\dagger}$          \\
MegaTTS 3         & 339M                  & 60kh EN                & \uline{$-0.02^\diamondsuit$} & $\textbf{2.31}$*  & $\textbf{0.70}$*  & -                         \\
F5-TTS     & 336M                  & 100kh Multi.           & $-0.19^{\dagger}$            & \uline{$2.42$}*   & $0.66$*   & \uline{$3.96^\dagger$}      \\
\textbf{FlexSpeech\#50} & 88M+330M              & 100kh Multi.           & $-0.03$                       & $2.64$            & $0.60$            & $\textbf{3.98}$           \\ 
\textbf{FlexSpeech\#1k} & 88M+330M              & 100kh Multi.           & $\textbf{0.00}$                       & $2.64$            & $0.59$            & $\textbf{3.98}$           \\ \hline
\multicolumn{7}{c}{\textbf{Seed-TTS \textit{test-en}}}                                                                                                                \\ \hline
GT                & -                     & -                      & $-0.20$                      & $2.06$            & $0.73$            & $3.94$                    \\
CosyVoice         & $\sim$300M            & 170kh Multi.           & $-0.16^{\dagger}$            & $3.39$*           & $0.64$*           & $3.59^{\dagger}$          \\
FireRedTTS        & $\sim$580M            & 248kh Multi.           & $-1.12^{\dagger}$            & $3.82$*           & $0.46$*           & $3.36^{\dagger}$          \\
MaskGCT           & 1048M                 & 100kh Multi.           & $-0.14^{\dagger}$            & $2.623$*          & $\textbf{0.717}$* & $3.81^{\dagger}$          \\
F5-TTS     & 336M                  & 100kh Multi.           & $-0.11^\dagger$        & \uline{$1.83$}*  & \uline{$0.67$}*   & $\textbf{3.93}^{\dagger}$ \\
\textbf{FlexSpeech\#50} & 88M+330M              & 100kh Multi.           & \uline{$-0.02$}              & $1.86$     & $0.61 $           & $3.91$           \\
\textbf{FlexSpeech\#1k} & 88M+330M              & 100kh Multi.           & $\textbf{0.00}$              & $\textbf{1.81}$     & $0.62 $           & \uline{$3.92$ }             \\ \hline
\multicolumn{7}{c}{\textbf{Seed-TTS \textit{test-zh}}}                                                                                                                \\ \hline
GT                & -                     & -                      & $-0.21$                      & $1.26$            & $0.76$            & $3.78 $                   \\
CosyVoice         & $\sim$300M            & 170kh Multi.           & $-0.29^{\dagger}$            & $3.10$*           & $0.75$*           & $3.63^{\dagger}$          \\
FireRedTTS        & $\sim$580M            & 248kh Multi.           & $-0.77^{\dagger}$            & $1.51$*   & $0.63$*           & $3.44^{\dagger}$          \\
MaskGCT           & 1048M                 & 100kh Multi.           & $-0.15^{\dagger}$            & $2.273$*          & $\textbf{0.774}$* & $3.80^{\dagger}$          \\
F5-TTS     & 336M                  & 100kh Multi.           & $-0.10^\dagger$        & $1.56$*           & \uline{$0.76$}*   & $3.85^\dagger$      \\
\textbf{FlexSpeech\#50} & 88M+330M              & 100kh Multi.           & \uline{$-0.03$}              & \uline{$1.29$}   & $0.68$            & \uline{$3.91$}           \\
\textbf{FlexSpeech\#1k} & 88M+330M              & 100kh Multi.           & $\textbf{0.00}$              & $\textbf{1.20}$   & $0.68$            & $\textbf{3.93}$           \\ \hline
\end{tabular}

\label{table_total}
\end{table*}

\subsection{Evaluation Results}
We compared our models with previous state-of-the-art (SOTA) zero-shot TTS systems including MaskGCT, NaturalSpeech 3, Mega-TTS 2, Mega-TTS 3, CosyVoice, FireRedTTS, and F5-TTS. As shown in Table~\ref{table_total}, our FlexSpeech achieves superior zero-shot TTS naturalness and expressiveness while maintaining strong stability.
For intelligibility, FlexSpeech achieves the lowest WER score of $1.20$ in Seed-TTS \textit{test-zh} and outperformed all previous SOTA models of $1.81$ in Seed-TTS \textit{test-en}. This indicates that the stability performance of FlexSpeech has achieved SOTA.
Regarding speaker similarity of timbre and prosody pattern (SMOS), we achieved superior SMOS scores compared to all baselines of $3.93$ in Seed-TTS \textit{test-zh} and $3.98$ in LibriSpeech-PC \textit{test-clean}. In Seed-TTS \textit{test-en}, we obtained the second-highest score $3.92$, comparable to that of F5-TTS. In terms of naturalness and expressiveness (CMOS), for non-open-source systems (NaturalSpeech 3, MegaTTS 2, MegaTTS 3), we download samples corresponding to the test set from demo pages, then infer FlexSpeech to obtain parallel samples for subjective evaluation. FlexSpeech performs slightly better than or comparable to theirs in these samples. However, compared to all other baselines, FlexSpeech demonstrates superiority in naturalness evaluations.
These results demonstrate that FlexSpeech has achieved state-of-the-art stability performance while also attaining top-tier naturalness and expressiveness, further validating the effectiveness of our decoupled framework.
Additionally, we observed that using 50 data pairs during the DPO phase yields roughly the comparable level of model enhancement to that of using 1000 data pairs. This suggests that DPO can achieve significant improvements for our duration model with just a few dozen preference data pairs, and further increases in data scale lead to minimal additional benefit.

\subsection{Ablation Study}
We explore the impact of supervised fine-tuning and DPO optimization on our model performance. 
Specifically, We conduct three ablation systems which ablating the preference optimization and SFT phase for the duration model, and the SFT phase for the duration model, respectively.
Results are reported in Table~\ref{ablationstudy}. When DPO is not applied, both the WER and subjective speaker similarity SMOS have performance degradation. This indicates that DPO contributes to enhancements in naturalness and stability for our duration model.
We speculate that the reason might be that the duration model tends to produce prosodic patterns that do not align with human preference. For instance, it will generate unnatural pauses or overly mechanical duration patterns, thereby compromising the intelligibility and naturalness of speech.
If the SFT phase for the duration model is also ablated, we observe a significant decrease in WER and SMOS scores on both Chinese and English test sets. We speculate that this is due to the fact that the pre-training data, Emilia, was sourced from the internet and processed through an auto-process pipeline, which may introduce discrepancies between audios and transcriptions.  This results in alignment errors in extracted durations, thereby impacting the capacity of the duration model. When the duration model predicts inappropriate or unstable duration patterns, it becomes challenging for the acoustic model to generate clear and natural speech.
Furthermore, if we do not perform SFT on the acoustic model, there is also a decline in all four scores.
We attribute this to similar reasons. Due to our completely decoupled framework, the acoustic model's input is the phoneme sequence that has been fully expanded with durations. If there is a bias between the phoneme sequence and the audio, it can confuse the acoustic model and consequently limit its capabilities.

\section{Rapid Style Transfer}
Despite utilizing only a few dozen win-lose data pairs for direct preference optimization, it has been proven to significantly enhance the stability and naturalness of FlexSpeech. Additionally, thanks to the high controllability of our framework, we can rapidly transfer any specific style with just a few hundred data pairs onto speakers. We conduct experiments on the open-source StoryTTS~\cite{storytts}~\footnote{https://github.com/X-LANCE/StoryTTS} dataset, which is a highly expressive storytelling TTS dataset from the recording of a Mandarin storytelling show that contains rich expressiveness both in acoustic and textual perspective, and accurate text transcriptions. 
Our goal is to effortlessly transfer this style to unseen speakers using a very small amount of data. So we randomly selected 100 sentences from it as the test set, and several sets of a certain number of samples that do not overlap with the test set were used as train set to construct data pairs.

\begin{figure}[htbp]
    \centering
    \includegraphics[width=\columnwidth]{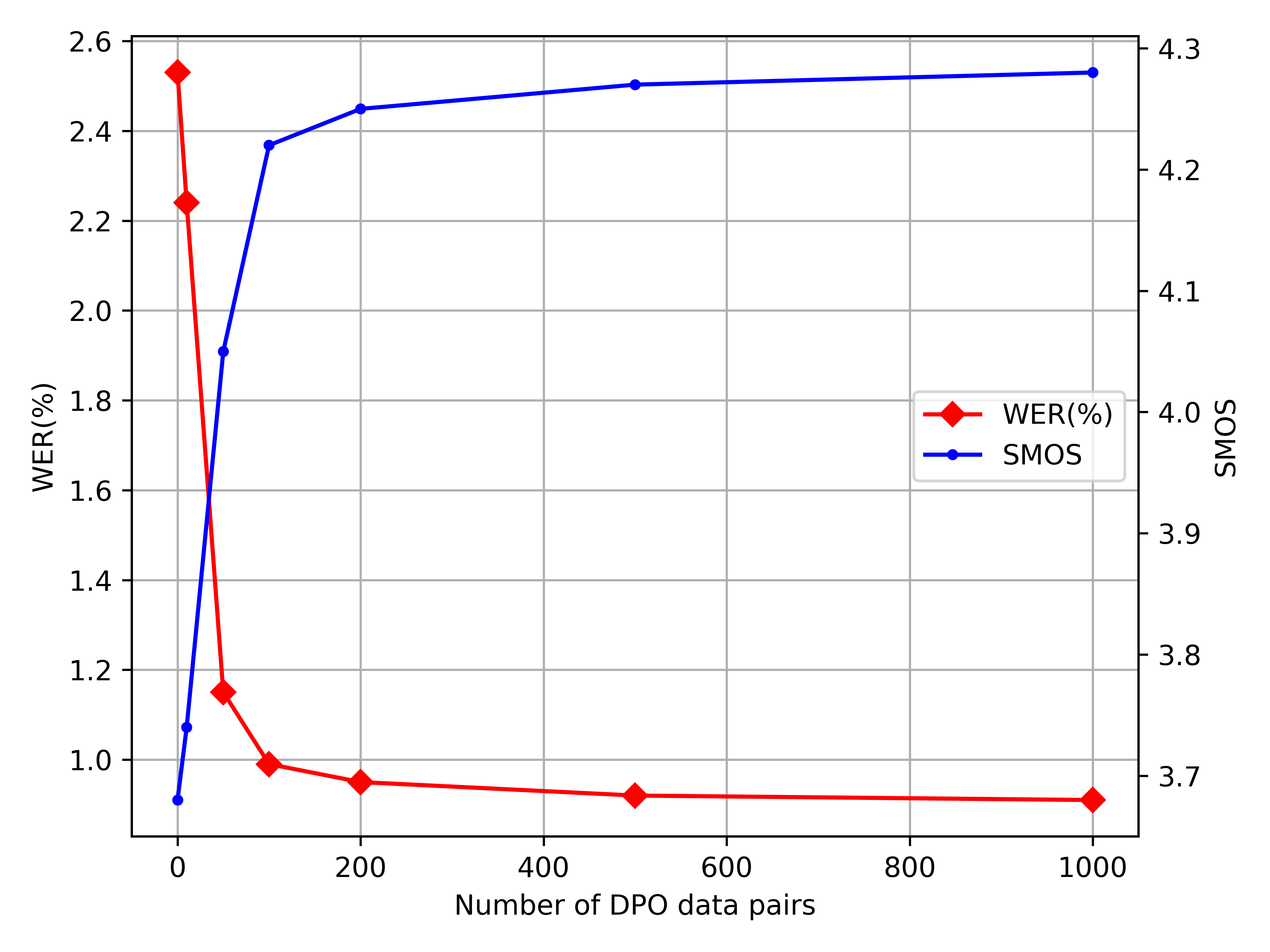}
    \caption{The WER and SMOS results of rapid style transfer. `0' on the horizontal axis means not applying DPO. `GT' means ground truth. }
    \label{styletransfer}
    \Description{The results of rapid style transfer. The red line shows the change of WER scores with the number of DPO data pairs. The blue line shows the change of SMOS scores with the number of DPO data pairs.}
\end{figure}

Specifically, We randomly select samples from the dataset as prompts for the duration model and predict the phoneme-level durations of the train set samples. These predicted durations served as negative examples, while the ground truth durations are considered positive examples, forming winner-loser data pairs for DPO.
We apply DPO on the duration model after SFT, using 10, 50, 100, 200, 500, and 1000 data pairs respectively, and evaluated it on the same test set. The predicted durations of test set are fed into the acoustic model to synthesize speech, with target speakers randomly selected from the Seed-TTS \textit{test-zh}. The results for WER and SMOS are illustrated in Figure~\ref{styletransfer}.
The results of the analysis indicate that WER scores exhibit a decreasing trend with the increase in the number of preference data pairs, stabilizing at a value of 50, beyond which further increases in data quantity do not yield significant benefits. SMOS scores demonstrate an upward trend as the number of DPO data pairs increases, reaching a plateau at a count of 100.
The results demonstrate that only a minimal amount of approximately 100 data pairs is sufficient for rapidly transferring a specific style to other speakers through performing direct preference optimization on the duration model of FlexSpeech, without necessitating any adjustments to the acoustic model.

\begin{table}[b]
\caption{Results of ablation study. `w/o DM-DPO,SFT' means without applying supervised fine-tuning and direct preference optimization to the duration model. `w/o AM-SFT' means without applying supervised fine-tuning to the acoustic model.}
\begin{tabular}{c|cc|cc}
\toprule
\multirow{2}{*}{\textbf{Method}} & \multicolumn{2}{c|}{\textbf{Seed-TTS \textit{test-en}}} & \multicolumn{2}{c}{\textbf{Seed-TTS \textit{test-zh}}} \\ \cline{2-5} 
                        & WER(\%)↓            & SMOS            & WER(\%)↓            & SMOS           \\ \midrule
FlexSpeech              & {$\textbf{1.86}$}               & $\textbf{3.91}$            & $\textbf{1.20}$               & $\textbf{3.89}$           \\ \hline
w/o DM-DPO              & \uline{$2.94$}                & $3.71$            & \uline{$2.54$}                & $3.72$           \\
w/o DM-DPO,SFT          & $6.65$                & $3.31$            & $5.93$                & $3.42$           \\ \hline
w/o AM-SFT              & $3.44$                & \uline{$3.88$}            & $3.17$                & \uline{$3.80$}           \\ \bottomrule
\end{tabular}
\label{ablationstudy}
\end{table}

\section{Related Work}

\subsection{Autoregressive TTS}
In recent years, autoregressive generation techniques have made significant progress in the field of speech synthesis. Research in this area is generally categorized into two main approaches based on the prediction target: discrete representations and continuous representations.
Discrete representation methods model continuous speech signals by quantizing them into discrete tokens. Some studies adopt multi-layer codebooks as prediction targets, leveraging multiple codebooks to capture both the diversity and fine-grained features of speech signals. For example, the Valle series~\cite{valle,valle2} and SPEAR TTS~\cite{spear} utilize multi-level cascade models to sequentially predict different layers of codebooks, while Fish-Speech~\cite{fishspeech} and UniAudio~\cite{uniaudio} employ carefully designed prediction strategies to predict all layers simultaneously. Other works focus on using a single codebook as the target in order to simplify the model structure while preserving as much speech information as possible. In these cases, Tortoise TTS~\cite{tortoise}, SeedTTS~\cite{seedtts}, and BaseTTS~\cite{BaseTTS} use the token with constraining audio reconstruction by discretizing acoustic features as targets, whereas the CosyVoice~\cite{cosyvoice,cosyvoice2} directly employs ASR-supervised discrete speech as the prediction target and subsequently reconstructs audio through flow matching.
 In contrast, continuous representation methods directly describe speech signals using continuous features. Early approaches, such as Tacotron~\cite{tacotron,tacotron2}, achieved speech synthesis by using an RNN to predict Mel-spectrograms on a frame-by-frame basis. Melle~\cite{melle} also treats Mel-spectrograms as the prediction target, using latent sampling to generate the next frame. Another method~\cite{aear} extracts low-dimensional representations via an autoencoder (AE), and Kalle~\cite{kalle} employs a variational autoencoder (VAE) to predict the next probability distribution. Compared with discrete representations, continuous representations are better at capturing the fine details of speech and enhancing the diversity of the generated output.

\subsection{Non-Autoregressive TTS}
Non-autoregressive text-to-speech accelerates synthesis speed through parallel generation mechanisms. Early approaches, such as FastSpeech ~\cite{fastspeech2} and DelightfulTTS~\cite{delight}, relied on external alignment tools to obtain phoneme durations and achieved speech synthesis through the joint training of a duration predictor. Later, models like VITS~\cite{vits} and GradTTS~\cite{gradtts} employed MAS to realize unsupervised duration alignment, thereby streamlining the training process. More recent efforts have sought to extend zero-shot voice cloning within the non-autoregressive framework; for example, NaturalSpeech 2~\cite{ns2} incorporates latent diffusion model during decoding to further model speaker timbre, while Mega TTS2~\cite{mega2} utilizes prosody latent language model to effectively capture the prosody of the reference audio. These methods explicitly model the duration information of speech to enhance the stability of the generation process, although their overall expressiveness remains relatively modest. In contrast, another category of methods does not depend on alignment information—examples include diffusion model-based approaches such as E2 TTS~\cite{e2tts} and F5 TTS~\cite{f5tts}, as well as MaskGCT~\cite{maskgct}, which employs a MaskGit~\cite{maskgit} strategy to directly generate speech sequences without requiring additional alignment information. Although this alignment-free approach offers advantages in improving speech expressiveness, it simultaneously introduces challenges related to generation stability.

\subsection{Preference alignment in Speech Synthesis}

In the field of speech synthesis, numerous studies have explored incorporating human evaluations into language model-based TTS optimization to enhance the naturalness and expressiveness of generated speech. For instance, SpeechAlign~\cite{SpeechAlign} introduced the first DPO-based method, with the core idea of treating basic facts as preferred samples and generated outputs as non-preferred ones, thereby steering the model toward producing speech that aligns better with human expectations. Meanwhile, UNO~\cite{UNO} leverages unpaired preference data by accounting for the uncertainty inherent in subjective evaluation annotations, and RIO~\cite{RIO} adopts a Bayesian-inspired inverse preference data selection strategy to more precisely screen and utilize preference data. Additionally, further research~\cite{Preference_ailmt} has focused on filtering preference data across multiple evaluation dimensions, aiming to improve TTS generation quality in various aspects.

\section{Conclusions}
In this paper, we present FlexSpeech, a stable, controllable, and expressive zero-shot TTS system that leverages two decoupling model components to align text-speech and generate speech. 
The duration model predicts phoneme durations that are consistent with the prompt patterns through in-context learning. The flow-matching-based TTS model then takes the phoneme sequence expanded by the predicted durations as input and predicts mel-spectrogram of the same length. A BigVGAN vocoder is employed to convert mel-spectrograms back into 48kHz high-fidelity waveforms.
By integrating autoregressive duration modeling with non-autoregressive speech generation paradigms, FlexSpeech achieves rich prosody and expressive effects while maintaining strong stability.
Our experiments demonstrate that FlexSpeech outperforms several strong zero-shot TTS systems on intelligibility, speaker similarity, and naturalness.
Ablation studies demonstrate that supervised fine-tuning phase and direct preference optimization contribute to enhanced stability and naturalness for FlexSpeech.
Moreover, when transferring to a specific stylistic domain, we can accomplish light-weight optimization of the duration model solely with about 100 data samples, without the need to adjust the acoustic model, thereby enabling rapid and stable style transfer.

\bibliographystyle{ACM-Reference-Format}
\bibliography{sample-base}
\appendix
\section{Details For Preference Annotation}
Initially, we utilize automated techniques to pre-select the data. In this stage, samples are filtered by calculating the Word Error Rate (WER) and detecting abnormal pauses, which allows us to eliminate entries with excessive errors or irregular pausing. Following this automated pre-filtering, the remaining samples are then forwarded to listeners for manual annotation.
\begin{figure}[htbp]
    \centering
    \includegraphics[width=\linewidth]{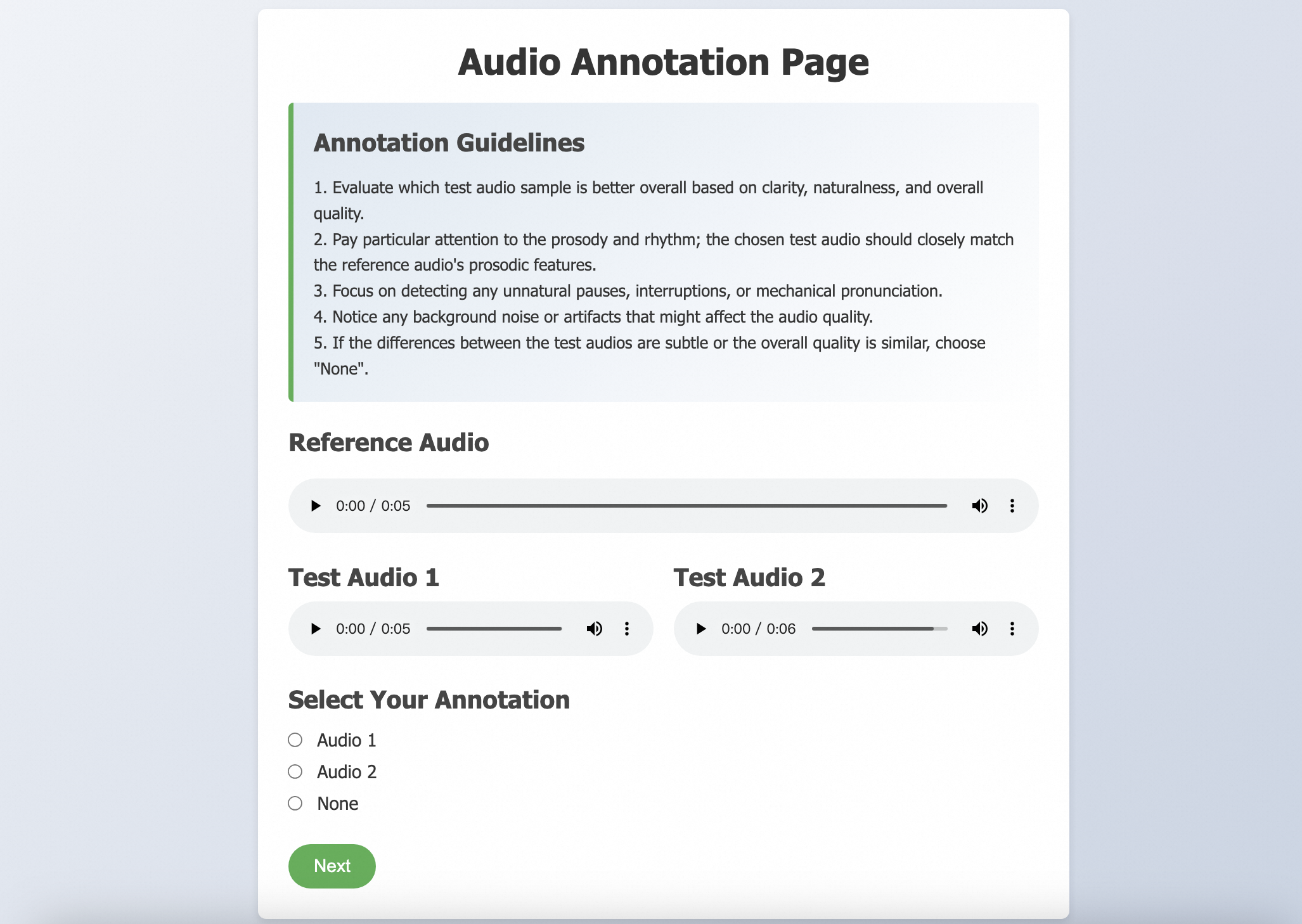}
    \caption{The interface of the annotation system.}
    \label{interface}
\end{figure}

\section{Duration Control}
\begin{figure}[h]
    \centering
    \includegraphics[width=\linewidth]{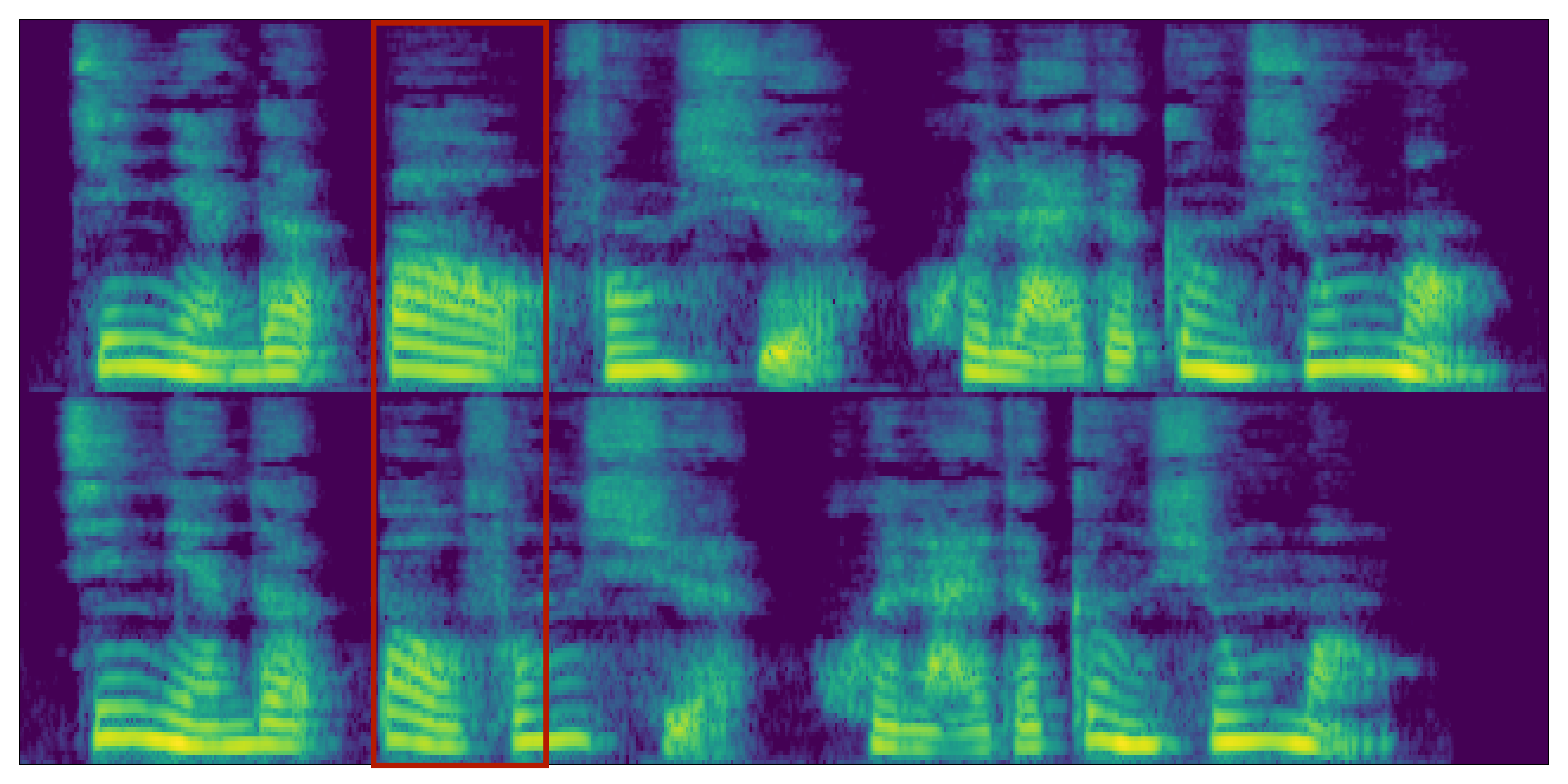}
    \caption{Phoneme-level duration control}
    \label{phoneme}
    \Description{The phoneme-level duration control example.}
\end{figure}

\begin{figure}[h]
    \centering
    \includegraphics[width=\linewidth]{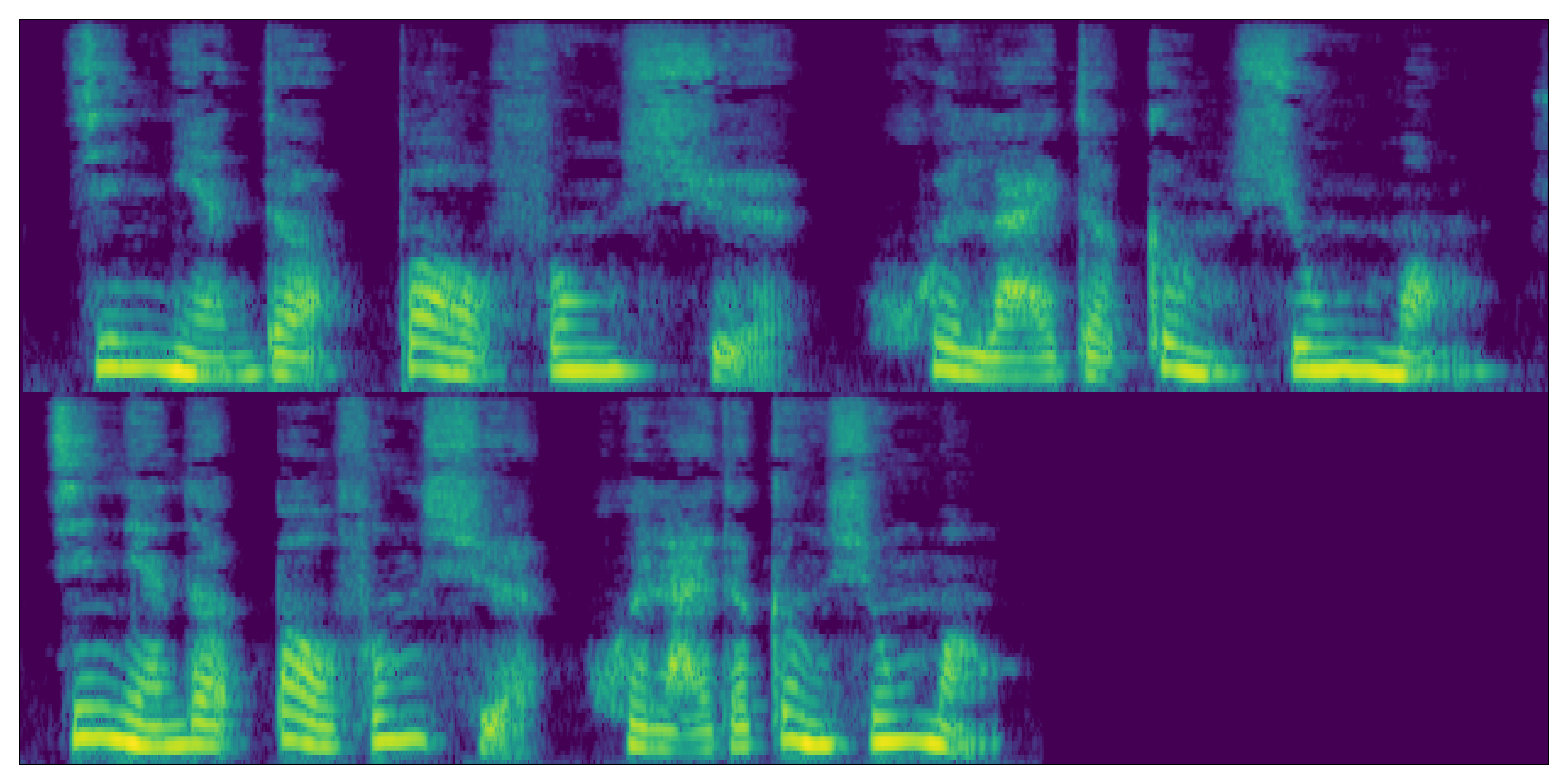}
    \caption{Sentence-level duration control}
    \label{sentence}
    \Description{The sentence-level duration control example.}
\end{figure}

We use a case study to demonstrate FlexSpeech's fine-grained duration control capability. We randomly selected a sample from the test set as a duration prompt, predicting the phoneme-level duration of the text `This year's snowstorm will be more fierce'. As shown in Figure~\ref{phoneme}, we perturb one of the phonemes by multiplying its duration by a coefficient of 1.5, resulting in audio output with doubled duration for that specific phoneme which is highlighted by the red box.  Then we apply the 1.5 multiplication factor to the durations of all phonemes in the sentence, slowing down the speech rate to 1.5 times its original pace, as shown in Figure~\ref{sentence}.

\end{document}